\documentstyle[12pt]{article}

\newcommand{\sect}[1]{\setcounter{equation}{0}\section{#1}}

\newcommand{\bea}{\begin{eqnarray}}
\newcommand{\eea}{\end{eqnarray}}
\newcommand{\be}{\begin{equation}}
\newcommand{\ee}{\end{equation}}
\newcommand{\vs}[1]{\vspace{#1 mm}}

\newcommand{\dsl}{\pa \kern-0.5em /}

\newcommand{\pa}{\partial}

\newcommand{\nn}{\nonumber\\}

\newcommand{\ddA}{\ddot{A}}
\newcommand{\dA}{\dot{A}}
\newcommand{\dB}{\dot{B}}
\newcommand{\ddB}{\ddot{B}}
\newcommand{\ddG}{\ddot{G}}
\newcommand{\dG}{\dot{G}}

\begin{document}
\topmargin 0pt
\oddsidemargin 0mm

\begin{flushright}
hep-th/0309202\\
\end{flushright}

\vs{2}
\begin{center}
{\Large \bf  
Time dependent supergravity solutions in arbitrary dimensions}
\vs{10}

{\large Somdatta Bhattacharya and Shibaji Roy}
\vspace{5mm}

{\em 
 Saha Institute of Nuclear Physics,
 1/AF Bidhannagar, Calcutta-700 064, India\\
E-Mails: som, roy@theory.saha.ernet.in\\}
\end{center}

\vs{10}
\centerline{{\bf{Abstract}}}
\vs{5}
\begin{small}
By directly solving the equations of motion we obtain the time dependent
solutions of supergravities with dilaton and a $q$-form field strength
in arbitrary dimensions. The metrics are assumed to have the symmetries
ISO($p+1$) $\times$ SO($d-p-2,1$) and can be regarded as those of 
the magnetically charged
Euclidean or space-like branes. When we impose the extremality condition,
we find that the magnetic charges of the branes become imaginary and the
corresponding real solutions then represent the E$p$-branes of 
type II$^\ast$ theories
(for the field-strengths belonging to the RR sector).
On the other hand, when the extremality condition is relaxed we find  real
solutions in type II theories which resemble the solutions found by 
Kruczenski-Myers-Peet.
In $d=10$ they match exactly. We point out the relations between the
solutions found in this paper and those of Chen-Gal'tsov-Gutperle
in arbitrary dimensions. Although there is no extremal limit for these
solutions, we find another class of solutions, which resemble the solutions
in the extremal case with imaginary magnetic charges and the corresponding
real solutions can be regarded
as the non-BPS E$p$-brane solutions of type II$^\ast$ theories (for the 
field-strengths in RR sector).
\end{small}

\newpage

\sect{Introduction}

Low energy effective actions of (dimensionally reduced) string/M theories
are known \cite{lmpx,im,bqr,kosst} to admit various kinds of time 
dependent solutions. These 
solutions are interesting from the cosmological point of view 
\cite{bqr,kosst} and are 
believed to shed light on the dS/CFT correspondence \cite{as,dk,bdm,mp}. 
One particular class
of solutions which might be useful in this context 
were assumed in the literature to have the symmetry ISO($p+1$) $\times$
SO($d-p-2,1$) of the metric in $d$ space-time dimensions and can be
regarded as those of magnetically charged Euclidean or Space-like branes
\cite{gs} 
(S$p$-branes\footnote{S$p$-branes by definition have $(p+1)$-dimensional
Euclidean world-volume.}
for short). These solutions were found by Kruczenski-Myers-Peet (KMP) in
$d=10$ \cite{kmp} and by Chen-Gal'tsov-Gutperle (CGG) in arbitrary $d$ 
\cite{cgg} in two
different coordinate systems. Though in $d=10$, these two solutions can 
be shown \cite{srone} to be equivalent by a coordinate transformation, 
the two coordinate 
systems have the advantages of their own. So, for example, the solutions
in CGG coordinates were found to be useful to obtain a four-dimensional
FRW universe with accelerating cosmology \cite{no,srtwo}, whereas, 
the KMP coordinates are
more suitable to understand the dS/CFT correspondence.

In \cite{kmp}, KMP obtained the supergravity S-brane solutions in $d=10$,
starting from the known solutions \cite{cmp} of vacuum Einstein equation 
in $d=11$ with 
appropriate symmetries and using a rotation, compactification to $d=10$
as well as applying T-duality symmetries \cite{bho,bmm}. This way one 
could avoid solving
the non-linear differential equations resulting from Einstein's equations.
In this paper, we directly solve the non-linear equations of motion (in KMP
coordinates) of the supergravities containing the dilaton and an abitrary
rank field-strength to obtain the S-brane solutions in $d$ dimensions.
We assume the metrics to have the symmetries ISO($p+1$) $\times$ SO($d-p-2,1$)
and therefore obtain only the localized and isotropic (as opposed to 
ref.\cite{kmp}, where anisotropic S-brane solutions were also found) S-brane
solutions. One advantage of this direct method is that one can see what
are the other kinds of solutions the equations of motion admit. In fact,
we find that if the extremality condition (similar to the static BPS $p$-brane
solutions \cite{dkl}) is assumed to hold, the magnetic charges of the 
branes become 
imaginary. This implies that the corresponding real solutions represent 
E$p$-branes of type
II$^\ast$ theories when the field-strengths belong to RR sector. The `starred'
theories were introduced by Hull \cite{ch} in studying the compactifications
of type II string theories on a time-like circle\footnote{The actions for the
`starred' theories can be obtained from those of the orinary theories by
changing the signs of the kinetic terms of the RR sector gauge fields,
but the kinetic terms of the NSNS sector gauge fields do not change 
sign.}. Actually, it was shown
there that type IIA (IIB) string theory compactified on a time-like circle
of radius $R$ is equivalent to type IIB$^\ast$ (IIA$^\ast$) string theory
compactified on a dual time-like circle of radius $1/R$.  
The `starred' theories
have a number of problems as emphasized by Hull in \cite{ch}. However, if
time-like T-duality is consistent in string theories, then `starred' theories
are also consistent as string theories. It was pointed out in \cite{ch} that
`starred' string theories truncated to the supergravity level may give
rise to theories with ghosts, but the full string theories would be ghost-free
as they are equivalent to the ordinary type II theories by a T-duality
(time-like) transformation.
On the other hand, if the extremality condition is relaxed, then
real solutions can be found in type II theories and the equations of motion 
can be solved
consistently if we introduce at least three unknown parameters\footnote{Here
we also restrict our metrics to become asymptotically flat (in Rindler
coordinates) and the dilaton to approach unity. If we do not impose this 
restriction, then our solutions will be characterized by an additional 
parameter.}. These solutions resemble the solutions obtained by KMP in
\cite{kmp}. In $d=10$ they match exactly with the KMP solutions for the
isotropic case. We also point out how these solutions in arbitrary $d$
are related to the solutions found by CGG in ref.\cite{cgg} by a coordinate
transformation. We note that there is no extremal limit for these solutions.
However, there exist another class of solutions for which the magnetic charges
are imaginary and resemble the solutions in the extremal case and we point
out that the corresponding real solutions represent the non-BPS E$p$-branes 
of type II$^\ast$
theories again for the field-strengths in RR sector.

This paper is organized as follows. In section 2, we obtain the time
dependent solution of the supergravity equations of motion with the
metrics having the symmetries ISO($p+1$) $\times$ SO($d-p-2,1$) and the 
components satisfying the extremality condition. Here we obtain the
Euclidean brane solutions with imaginary magnetic charges, representing the
E$p$-branes of type II$^\ast$ theories. In section 3, we relax the
extremality condition and obtain real time dependent solutions 
in type II theories characterized
by three parameters. In subsection 3.1, we show that for $d=10$, our
solutions match exactly with those obtained by KMP in \cite{kmp}. In subsection
3.2, we show how for arbitrary $d$ our solutions are related to those
of CGG by a coordinate transformation similar to that found in \cite{srone}.
In section 4, we discuss another class of solution which is non-extremal
but is very similar to the extremal solutions found in section 2 and the
magnetic charges are imaginary. We point out that these solutions can be 
interpreted as the non-BPS E$p$-brane solutions of type II$^\ast$ theories. We
conclude our paper in section 5.

\sect{General extremal Euclidean brane solutions}

In this section we discuss the Euclidean brane\footnote{Euclidean branes
in type II theories are termed as S-branes, whereas Euclidean branes in type
II$^\ast$ theories are termed as E-branes for the field-strengths belonging
to RR sector.} solutions of supergravity equations of motion in $d$-dimensions
when the metric components satisfy an extremality condition similar to the
static BPS $p$-brane solutions. The $d$-dimensional action of a graviton,
dilaton and a $q$-form field-strength with dilaton coupling $a$ in Einstein 
frame has the form,
\be
S = \int d^dx \sqrt{-g} \left[R - \frac{1}{2} \partial_\mu \phi \partial^\mu
\phi - \frac{1}{2\cdot q!} e^{a\phi} F_{[q]}^2\right]
\ee
The above action is quite general and consists of the bosonic sector of
(dimensionally reduced) string/M theories. The field-strength in (2.1) is
real, but as pointed out in \cite{ch}, it could be imaginary if, for
example, it belongs to RR sector of type II theory. For the latter case,
the action could be obtained from the so-called `starred' theory with
the real field-strength but the kinetic term will have opposite sign from
that of ordinary theory. 

The equations of motion following from (2.1) have the forms,
\bea
R_{\mu\nu} - \frac{1}{2} \partial_\mu \phi\partial_\nu \phi - 
\frac{e^{a\phi}}
{2(q-1)!} \left[F_{\mu\alpha_2\ldots \alpha_q}F_\nu^{\,\,\,\alpha_2
\ldots \alpha_q}
- \frac{q-1}{q(d-2)}F^2_{[q]}g_{\mu\nu}\right] &=& 0\\
\partial_\mu\left(\sqrt{-g}e^{a\phi} F^{\mu\alpha_2\ldots 
\alpha_q}\right) &=& 0\\ 
\frac{1}{\sqrt{-g}}\partial_\mu\left(\sqrt{-g}\partial^\mu \phi\right)
- \frac{a}{2 \cdot q!} e^{a\phi} F_{[q]}^2 &=& 0
\eea
We will solve the above equations with the following ansatz,
\bea
ds^2 &=& e^{2A(t)}\left(-dt^2 + t^2 dH_{d-p-2}^2\right) + e^{2B(t)} \left(
dx_1^2 + \cdots + dx_{p+1}^2\right)\\
F_{[q]} &=& b\,\, {\rm Vol}(H_{d-p-2})
\eea
Here `$t$' is a time-like coordinate and $A$, $B$ are functions of `$t$'
whose forms will be obtained by solving the equations of motion (2.2) --
(2.4). $dH^2_{d-p-2}$ is the line element of a unit $(d-p-2)$ dimensional
hyperbolic space and Vol($H_{d-p-2}$) is its volume-form. $q=d-p-2$ is the
rank of the field-strength and $b$ is the magnetic charge. Note that the metric
in (2.5) has the symmetry ISO($p+1$) $\times$ SO($d-p-2,1$). The Ricci tensor
of the hyperbolic space is given as $\bar{R}_{ab} = -(q-1) \bar{g}_{ab}$,
with $\bar{g}_{ab}$ being its metric. Also the metric has the form of an
Euclidean brane with $(p+1)$-dimensional world-volume whose transverse
space metric, upto a conformal factor $e^{2A(t)}$, can be written as,
\bea
-dt^2 + t^2 dH_{d-p-2}^2 &=& -dt^2 + t^2 d\psi^2 + t^2 \sinh^2\psi 
d\Omega_{d-p-3}^2\nonumber\\
&=& -d\tilde{t}^2 + dr^2 + r^2 d\Omega_{d-p-3}^2
\eea
with $d\Omega_{d-p-3}^2$ being the line element of unit $(d-p-3)$-dimensional 
sphere. In the last expression of (2.7) we have used $t^2 = \tilde{t}^2 - r^2$
and $\tanh \psi = r/\tilde{t}$. This shows that the transverse space is flat
upto a conformal factor in $\tilde{t}$, $r$ coordinates and the metric 
has the required symmetry, since $e^{2A(t)}$ is invariant under SO($d-p-2,1$).
We will also assume that $A(t)$, $B(t)$ will vanish as $t \to \infty$, i.e.
the metric is asymptotically flat (in Rindler coordinates). So, (2.5) and
(2.6) represent the magnetically charged Euclidean branes. The corresponding
electrically charged branes can be obtained by using $g_{\mu\nu} \to 
g_{\mu\nu}$, $F \to e^{-a\phi} \ast F$, $\phi \to -\phi$, where $\ast$ is
the $d$-dimensional Hodge dual.

Since eq.(2.3) is satisfied with the ansatz (2.6), we will solve (2.2) using
the ansatz (2.5), (2.6). The non-zero components of the Ricci tensor are
given below,
\bea
R_{tt} &=& -(p+1) \left[\ddot{B} + \dot{B}^2 - \dot{A}\dot{B}\right] -
q\left[\ddot{A} + \frac{\dot{A}}{t}\right]\nonumber\\
R_{xx} &=& e^{2B-2A}\left[\ddot{B} + (q-1) \dot{A}\dot{B} + (p+1) \dot{B}^2
+ q \frac{\dot{B}}{t}\right]\nonumber\\
R_{ab} &=& t^2\left[\ddot{A} + (q-1) \dot{A}^2 + (2q-1) \frac{\dot{A}}{t}
+ (p+1) \dot{B} (\dot{A} + \frac{1}{t})\right]\bar{g}_{ab}
\eea
Now we note that if we use the extremality condition similar to the static
BPS $p$-brane,
\be
(p+1) B + (q-1) A = 0
\ee
then the components of the Ricci tensor given in (2.8) simplify to
\bea
R_{tt} &=& - \left(\ddot{A} + \frac{q}{t} \dot{A} + \frac{(q-1)(d-2)}{p+1}
\dot{A}^2\right)\nonumber\\ 
R_{xx} &=& e^{2B-2A}\left(\ddot{B} + \frac{q}{t}\dot{B}\right)\nonumber\\
R_{ab} &=& t^2\left(\ddot{A} + \frac{q}{t}\dot{A}\right)
\bar{g}_{ab}
\eea
Here `dot' represents the derivative with respect to $t$. 
Substituting $R_{xx}$,
$R_{ab}$, $R_{tt}$ in (2.2) and from the $\phi$ equation in (2.4) we get,
\bea
\ddB + \frac{q}{t}\dB + \frac{b^2(q-1)}{2(d-2)} \frac{e^{2(p+1)B+a\phi}}
{t^{2q}} &=& 0\\
\ddA + \frac{q}{t}\dA - \frac{b^2(p+1)}{2(d-2)} \frac{e^{2(p+1)B+a\phi}}
{t^{2q}} &=& 0\\
\ddA + \frac{q}{t}\dA + \frac{(q-1)(d-2)}{(p+1)} \dA^2 + \frac{1}{2}
\dot{\phi}^2 + \frac{b^2(q-1)}{2(d-2)} \frac{e^{2(p+1)B+a\phi}}
{t^{2q}} &=& 0\\
\ddot{\phi} + \frac{q}{t}\dot{\phi} + \frac{a b^2}{2} 
\frac{e^{2(p+1)B+a\phi}}{t^{2q}} &=& 0
\eea
Comparing $R_{xx}$ equation (2.11) and $\phi$ equation (2.14) we find,
\be
\phi = \frac{a(d-2)}{q-1} B
\ee
Now using the extremality relation (2.9), we note that $R_{xx}$ and $R_{ab}$
equations given in (2.11) and (2.12) are identical and take the form,
\be
\ddB + \frac{q}{t} \dB + \frac{b^2(q-1)}{2(d-2)} \frac{e^{B\chi}}{t^{2q}} = 0
\ee
where we have used eq.(2.15). Also, in the above $\chi = 2(p+1) + 
a^2(d-2)/(q-1)$. Now if $H(t)$ is a harmonic function\footnote{We would like
to point out that although we call this function as harmonic function (in 
analogy with static solutions) throughout this paper, it is not really a 
solution of $(q+1)$-dimensional Laplace equation in spherical coordinates. 
It is rather a solution of Laplace-like equation in hyperbolic coordinates,
where, there are one time-like and $q$ space-like coordinates.
Also $t$ above is not a radial coordinate. It
is a time-like coordinate with SO($d-p-2,1$) invariance as mentioned earlier.}
in the $(q+1) = 
(d-p-1)$-dimensional transverse space, then $H(t)$ satisfies,
\be
\ddot{H} + \frac{q}{t}\dot{H} = 0
\ee
Putting $e^B = H^\alpha$, where $\alpha$ is a constant to be determined from
eq.(2.16), this equation reduces to a first order differential equation of
$H$. Then $H$ is determined from eq.(2.16) as,
\be
H = \left[1+\sqrt{\frac{b^2}{2(q-1)(d-2)\alpha}}\frac{1}{t^{q-1}}\right]
\ee
and the constant $\alpha$ is given as, $\alpha = -2/\chi$. Now since $\chi
> 0$, we have $\alpha < 0$, therefore demanding the harmonic function $H$
to be real we find $b^2 < 0$. In other words, the $q$-form field-strength
$F_{[q]}$ given in (2.6) is purely imaginary. We therefore have,
\be
e^{2B} = H^{2\alpha} = H^{-4/\chi}, \qquad e^{2A} = e^{-\frac{2(p+1)}{q-1}B}
= H^{\frac{4(p+1)}{\chi(q-1)}}
\ee
The solutions then take the form,
\bea
ds^2 &=& H^{\frac{4(p+1)}{\chi(q-1)}}\left(-dt^2 + t^2 dH_{d-p-2}^2\right)
+ H^{-\frac{4}{\chi}}\left(dx_1^2 + \cdots + dx_{p+1}^2\right)\nn
F_{[q]} &=& b\,\,{\rm Vol}(H_q), \qquad e^{2\phi}\,\,=\,\, H^{-\frac{4a(d-2)}
{(q-1)\chi}}
\eea
with `$b$' being purely imaginary and $H$ is as given in (2.18). 
It can be easily 
checked that this solution satisfies $R_{tt}$ equation (2.13). The solutions 
(2.20) represent $(p+1)$-dimensional Euclidean branes with imaginary 
magnetic charges and the imaginary charge is a result of the 
extremality condition
(2.9). The solutions are given in the Einstein frame and have exactly the same 
form as the static BPS $p$-brane solutions of type II string theories
\cite{dkl}. For 
various values of $a$, $\chi$ and $q$ they represent different brane 
solutions of (dimensionally reduced) string/M theories. A list of these 
parameters for $d=11$, 10 is given in table I of ref.\cite{cgg}. The only case
we have to be careful is when $q=5$ in $d=10$. Because in that case, the
field-strength would be self-dual and is not considered in our solutions.
For this case the dilaton coupling $a$ would be zero and by self-duality
$F_{[5]}^2=0$. Therefore from the dilaton equation of motion (2.4), we see
that it can be set to a constant. The field-strength in that case would take
the form,
\be
F_{[5]} = \frac{b}{\sqrt{2}}(1+\ast)\,\, {\rm Vol}(H_5)
\ee
The metric can be obtained from (2.20) by putting $a=0$. We note that when
$F_{[q]}$'s belong to RR sector of string theories, the solutions (2.20) would
be real if we start from `starred' theory action. Then the solutions
(2.20) would represent the E$p$-branes of type II$^\ast$ theories. However,
when $F_{[q]}$'s belong to the NSNS sector, (2.20) would be real if the action
contains kinetic terms of the NSNS sector gauge fields with signs opposite
from those of the ordinary theories. But the relation between these
theories and the ordinary type II theories is not clear to us. 
Similar
situation arises also for $d=11$. In this case, $F_{[4]}^2$ term or $F_{[7]}^2$
term has opposite signs form that in ordinary M-theory. However, the 
dimensional reductions (along a space-like direction) of this theory leads
neither to type IIA theory nor to type IIA$^\ast$ theory. So, as in 
the previous
case the relation between this M-theory and ordinary M-theory or type 
IIA/IIA$^\ast$ theory is not clear to us.

\sect{S-brane solutions in arbitrary dimensions}

We have seen in the previous section that on imposing the extremality condition
(2.9), the solutions of the equations of motion (2.2) -- (2.4) become
imaginary\footnote{If we restrict ourselves only to type II theories this
would mean that for the time dependent case there are no extremal solutions.
However, here we are using the notion of extremality in the broader sense
including the `starred' theories. The function $G(t)$ defined below is
called the non-extremality function also in this broader sense.}. 
In this section, we will see that by relaxing the condition
(2.9), we can get real solutions. We modify the condition (2.9) as follows,
\be
(p+1) B + (q-1) A = \ln G(t)
\ee
Now using (3.1) we obtain the non-zero components of the Ricci tensor from
(2.8) as,
\bea
R_{tt} &=& -\ddot{A} - \frac{\ddG}{G} + \frac{\dG^2}{G^2} - \frac{1}{p+1}
\left(\frac{\dG}{G} - (q-1)\dA\right)^2 - (q-1) \dA^2 + \frac{\dG}{G}\dA 
- \frac{q}{t} \dA\\
R_{xx} &=& e^{2B-2A}\left(\ddot{B} + \frac{\dG}{G}\dB
+ \frac{q}{t}\dot{B}\right)\\
R_{ab} &=& t^2\left(\ddot{A} + \frac{q}{t}\dA + \frac{\dG}{G}(\dA + \frac{1}{t}
)\right)\bar{g}_{ab}
\eea
Substituting these in (2.2) and (2.4), the $R_{xx}$, $R_{ab}$, $R_{tt}$
and $\phi$ equations take the forms,
\bea
\ddB + \frac{q}{t}\dB + \frac{\dG}{G} \dB + \frac{b^2(q-1)}{2(d-2)} 
\frac{e^{2(p+1)B+a\phi}} {G^2 t^{2q}} &=& 0\\
\ddA + \frac{q}{t}\dA + \frac{\dG}{G} (\dA+\frac{1}{t}) - 
\frac{b^2(p+1)}{2(d-2)} \frac{e^{2(p+1)B+a\phi}} {G^2 t^{2q}} &=& 0\\
-\ddA - \frac{\ddG}{G} + \frac{\dG^2}{G^2} - \frac{1}{p+1}\left(\frac{\dG}{G}
- (q-1)\dA\right)^2 - (q-1) \dA^2 + \frac{\dG}{G} \dA - \frac{q}{t} \dA & & \nn
- \frac{1}{2} \dot{\phi}^2 
- \frac{b^2(q-1)}{2(d-2)} \frac{e^{2(p+1)B + a\phi}}{G^2 t^{2q}} &=& 0\\
\ddot{\phi} + \frac{q}{t}\dot{\phi} + \frac{\dG}{G}\dot{\phi} + 
\frac{a b^2}{2}\frac{e^{2(p+1)B+a\phi}}{G^2 t^{2q}} &=& 0
\eea
Using (3.1) into (3.6), we first convert this equation into an equation of 
the function $B(t)$ and then using (3.5) we find an equation involving the
function $G(t)$ only (this gives a restriction on the form of the 
non-extremality function) as,
\be
\ddG + \frac{2q-1}{t} \dG = 0
\ee
The solution of this equation is,
\be
G(t) = 1 \pm \frac{\omega^{2(q-1)}}{t^{2(q-1)}}
\ee
where $\omega$ is an integration constant. We have also assumed that as
$t \to \infty$, $G(t) \to 1$. However, note that in this limit there are no
real solutions (except the flat space) of the equations of motion as discussed
in the previous section. Since here we are discussing the real solutions,
there will not be any extremal limit for the solutions we obtain later.
We also point out that the upper sign in (3.10) does not lead to a
real solution
and we consider only the lower sign. Now $G(t)$ can be split up as,
\be
G(t) = 1 - \frac{\omega^{2(q-1)}}{t^{2(q-1)}} = \left(1-\frac{\omega^{q-1}}
{t^{q-1}}\right)\left(1+\frac{\omega^{q-1}}{t^{q-1}}\right) = H H'
\ee
where $H(t)$ and $H'(t)$ are two harmonic functions in the $(q+1)$-dimensional
transverse space satisfying equations of the form (2.17). We can now try 
to solve the equations for $B(t)$ in (3.5) in analogy with the previous 
section by choosing,
\be
B = \alpha \ln H - \beta \ln H'
\ee
where $\alpha$ and $\beta$ are two parameters to be determined from the 
equations of motion. Before using this form of $B$, we determine the form
of $\phi$ in terms of $B$ from equations (3.5) and (3.8). Using the last
two equations we obtain,
\be
\left(\ddot{\phi} - \frac{a(d-2)}{q-1}\ddB\right)
+ \frac{q}{t}\left(\dot{\phi} - \frac{a(d-2)}{q-1}\dB\right)
+ \frac{\dG}{G}\left(\dot{\phi} - \frac{a(d-2)}{q-1}\dB\right) = 0
\ee
The solution to this equation is given by,
\be
\phi = \frac{a(d-2)}{q-1} B + \delta \ln \frac{H}{H'}
\ee
where $\delta$ is an arbitrary constant and so,
\be
e^{2(p+1)B+a\phi} = \left(\frac{H}{H'}\right)^{a\delta} e^{B\chi}
\ee
with $\chi = 2(p+1) + a^2(d-2)/(q-1)$ as given before. Now using (3.15) and
(3.12) in (3.5) we see that this equation can be consistently solved if
\be
\alpha = \frac{1-a\delta}{\chi}, \qquad \beta = - \frac{1+a\delta}{\chi}
\ee
The solution then is,
\be
\omega^{2(q-1)} = \frac{b^2\chi}{4(d-2)(q-1)}
\ee
However, it can be easily checked that this solution does not satisfy the
$R_{tt}$ equation (3.7). The reason is we have not introduced enough 
number of parameters and the system of equations (3.5) -- (3.7) become
overdetermined. In order to get around this situation we will use a different
ansatz for $B$ (other than (3.12)) introducing more parameters such that
$R_{tt}$ equation (3.7) can be made consistent. So, we take the form of
$B$ as,
\be
e^B = \left[\cos^2\theta \left(\frac{H}{H'}\right)^\alpha + \sin^2\theta
\left(\frac{H'}{H}\right)^\beta\right]^\gamma = F^\gamma
\ee
where $F = \left[\cos^2\theta \left(\frac{H}{H'}\right)^\alpha + \sin^2\theta
\left(\frac{H'}{H}\right)^\beta\right]$. The justification for choosing 
this particular form of $B$ can be understood as follows. We have seen before
that $e^B = (H H')^\alpha (H')^{2a\delta/\chi}$ is inconsistent with the
equations of motion. So, for $\delta=0$, $e^B = (HH')^\alpha$ is inconsistent.
With the two harmonic functions $H$ and $H'$, we can form two other functions
i.e. $H/H'$ and $H'/H$. So, $e^B$ could be either $(H/H')^\alpha$ or 
$(H'/H)^\beta$, with $\alpha$, $\beta$ having the same sign or it could
be a linear combination of both of them. The linear combination would introduce
more parameters necessary for the consistency of the equations of motion.
Also, if we insist that as $t \to \infty$, $e^B \to 1$ then the only possible
linear combination is as given in (3.18). We have further raised this function
to the arbitrary power $\gamma$. Note that if we do not insist $e^B \to 1$
as $t \to \infty$ then $F$ could take the form $F = a_1^2 (H/H')^\alpha
+ a_2^2 (H'/H)^\beta$ and in that case instead of a single parameter $\theta$,
we introduce two parameters $a_1$ and $a_2$. Even in this case the equations of
motion can be solved consistently. However, for simplicity we choose the 
form of $e^B$ as in (3.18) and mention later how the solution would change
with the additional parameter.

Now using the form of $G$ (in eq.(3.11)), $\phi$ (in eq.(3.14)) and $B$ (in
eq.(3.18)) into eq.(3.5) we obtain,
\be
\left[4\gamma (q-1) \omega^{2(q-1)} (\alpha+\beta)^2 \sin^2\theta \cos^2\theta
\right]\frac{H^{\alpha-\beta-2} H'^{\beta-\alpha-2}}{F^2}
+\frac{b^2}{2(d-2)} H^{a\delta-2} H'^{-a\delta-2}F^{\gamma\chi} = 0
\ee
We thus obtain from here
\bea
\gamma \chi &=& -2, \qquad\qquad \alpha-\beta \,\,=\,\, a\delta\nn
\omega^{2(q-1)} &=& \frac{b^2\chi}{16(d-2)(q-1)(\alpha+\beta)^2\sin^2\theta
\cos^2\theta}
\eea
We thus get a consistent real solution of the equations of motion (3.5), (3.6)
and (3.8) in terms of four parameters $\alpha$, $\delta$, $\omega$ and 
$\theta$. Next we need to check whether this solution is consistent with
$R_{tt}$ equation (3.7). Substituting this solution into (3.7) we get for 
consistency a relation between the parameters as,
\be
\frac{1}{2}\delta^2 + \frac{2\alpha(\alpha-a\delta)(d-2)}{\chi(q-1)} =
\frac{q}{q-1}
\ee
Since using (3.21) we can eliminate one of the two parameters $\alpha$ 
or $\delta$ we therefore have real solutions with three independent
parameters $\alpha$, $\omega$ and $\theta$, very similar to the solutions
obtained by KMP in $d=10$ \cite{kmp}. So, the complete isotropic and localized 
S-brane solutions in $d$-dimensions
can be written as,
\bea
ds^2 &=& F^{\frac{4(p+1)}{(q-1)\chi}} (HH')^{\frac{2}{q-1}}\left(-dt^2 + t^2
dH_{d-p-2}^2\right) + F^{-\frac{4}{\chi}}\left(dx_1^2 + \cdots + 
dx_{p+1}^2\right)\nn
e^{2\phi} &=& F^{-\frac{4a(d-2)}{(q-1)\chi}} 
\left(\frac{H}{H'}\right)^{2\delta}\nn
F_{[q]} &=& b\,\,{\rm Vol}(H_{d-p-2})
\eea
where $F$ is given in eq.(3.18), $H$, $H'$ are given in eq.(3.11) with
$\omega$ given in (3.20). These solutions are characterized by four parameters
$\alpha$, $\delta$, $\omega$ and $\theta$ with a relation between $\alpha$
and $\delta$ given in (3.21). If instead of taking $e^B$ of the form (3.18),
we had taken 
\be
e^B = \left[a_1^2 \left(\frac{H}{H'}\right)^\alpha + a_2^2
\left(\frac{H'}{H}\right)^\beta\right]^\gamma = F^\gamma
\ee
then the solutions will be characterized by an additional 
parameter and $\omega$ would be given as,
\be
\omega^{2(q-1)} = \frac{b^2\chi}{16(d-2)(q-1)(\alpha+\beta)^2 a_1^2
a_2^2}
\ee
The solutions would then take exactly the same form as given in (3.22). 
Also, we would like to point out that in $d=10$ and for $q=5$, the
field-strength is self-dual. This case is not included in our previous 
discussion and the equations of motion in this case need to be solved 
separately. However, the solutions can be obtained from (3.22) by setting
$a=0$ with the 5-form field-strength taking the form as given earlier
in eq.(2.21) of section 2. Note that for $d=11$, (3.22) would represent
M-theory S-branes and in that case $\phi=0$. So, from (3.14) we find that
$a=0$ and $\delta=0$. When $\delta=0$, $\alpha=\beta$ and eq.(3.21)
will determine the value of the parameter $\alpha$. M-theory S-branes
will therefore have only two parameters $\omega$ and $\theta$. But because 
of the mismatch of the number of parameters (string theory branes have three
whereas M-theory branes have two parameters), the dimensional reductions
of M-theory S-branes will not reproduce string theory S-branes contrary to our
expectations. We have pointed out in ref.\cite{srthree}, that the isotropic and 
localized string theory S-branes can be obtained only from the delocalized 
(anisotropic) SM-branes by direct (double) dimensional reductions. 

\subsection{Relation with the KMP solution}

In this subsection we will show that the solutions (3.22) match exactly
with the KMP solutions \cite{kmp} in $d=10$. In $d=10$, $a=(p-3)/2$, $q=8-p$, 
$\chi=32/(7-p)$ for the space-like D$p$-branes and so, the solutions 
(3.22) take the forms,
\bea
ds^2 &=& F^{\frac{p+1}{8}} (HH')^{\frac{2}{7-p}}\left(-dt^2 + t^2
dH_{8-p}^2\right) + F^{-\frac{7-p}{8}}\left(dx_1^2 + \cdots + 
dx_{p+1}^2\right)\nn
e^{2\phi} &=& F^{\frac{3-p}{2}} 
\left(\frac{H}{H'}\right)^{2\delta}\nn
F_{[q]} &=& b\,\,{\rm Vol}(H_{8-p})
\eea
with $H=1-\omega^{7-p}/t^{7-p}$, $H'=1+\omega^{7-p}/t^{7-p}$ and $F$ is as 
given in eq.(3.18). Also,
\be
\omega^{2(7-p)} = \frac{b^2}{4(7-p)^2 (\alpha+\beta)^2\sin^2
\theta\cos^2\theta}
\ee
The parameters satisfy the relation,
\be
\frac{1}{2}\delta^2 + \frac{\alpha\beta}{2} = \frac{8-p}{7-p}
\ee
where $\delta$ is given by $\alpha - \beta = a \delta$. Since the KMP metric
is given in the string frame, we rewrite the metric in (3.25) also in string
frame as,
\be
ds^2 = F^{\frac{1}{2}}\left(\frac{H}{H'}\right)^{\frac{\delta}{2}}
(HH')^{\frac{2}{7-p}}\left(-dt^2 + t^2
dH_{8-p}^2\right) + F^{-\frac{1}{2}}\left(\frac{H}{H'}\right)^{\frac
{\delta}{2}}\left(dx_1^2 + \cdots + 
dx_{p+1}^2\right)
\ee
However, we note that this form of the metric is not quite the same as given in
eq.(17) of ref.\cite{kmp}. This is because the function $F$ we have defined in
(3.18) is not the same as $F_{\rm KMP}$ (see eq.(12) of ref.\cite{kmp}) for
the isotropic case. But we note that by defining
\bea
\alpha &=& \frac{3n(p-3)}{2(7-p)} - \frac{m}{2}\nn
\beta &=& -\frac{3n(p-3)}{2(7-p)} - \frac{m}{2}
\eea
where $m$, $n$ are two parameters used in \cite{kmp}, we can write,
\bea
F &=& \cos^2\theta \left(\frac{H}{H'}\right)^\alpha + \sin^2\theta \left(\frac
{H'}{H}\right)^\beta\nn
&=& F_{\rm KMP} \left(\frac{H}{H'}\right)^{\frac{2(p-4)n}{7-p}}
\eea
Substituting (3.30) in (3.28) we find that the string-frame metric takes the
form,
\be
ds^2 = F_{\rm KMP}^{\frac{1}{2}}\left(\frac{H}{H'}\right)^{n\frac{p-1}{7-p}}
(HH')^{\frac{2}{7-p}}\left(-dt^2 + t^2
dH_{8-p}^2\right) + F_{\rm KMP}^{-\frac{1}{2}}\left(\frac{H}{H'}\right)^n
\left(dx_1^2 + \cdots + 
dx_{p+1}^2\right)
\ee
where we have used $\alpha-\beta=6n(p-3)/(2(7-p))=a\delta$, with 
$\delta=6n/(7-p)$ and the dilaton takes the form,
\be
e^{2\phi} = F^{\frac{3-p}{2}}\left(\frac{H}{H'}\right)^{2\delta} = 
F_{\rm KMP}^{\frac{3-p}{2}} \left(\frac{H}{H'}\right)^{pn}
\ee
The parameter relation (3.27) now reduces to
\be
9n^2(p+1)+m^2(7-p)=8(8-p)
\ee
This is precisely the same form of the isotropic and localized space-like 
D$p$-brane solutions in $d=10$ obtained in ref.\cite{kmp}.

Now we show that the isotropic space-like NS-branes obtained in section
2.3 of ref.\cite{kmp} also match with the solutions given in (3.22). We
make an explicit comparison for the space-like NS5-brane (since this is
magnetically charged) and then mention how we compare the space-like 
NS1-brane solution. For the NS5-brane we have $d=10$, $p=5$, $a=-1$, $q=3$
and $\chi=16$. The solution (3.22) therefore takes the form in string-frame,
\bea
ds^2 &=& F (HH')\left(\frac{H}{H'}\right)^{\frac{\delta}{2}}\left(-dt^2 + t^2
dH_3^2\right) + \left(\frac{H}{H'}\right)^{\frac{\delta}{2}}
\left(dx_1^2 + \cdots + 
dx_6^2\right)\nn
e^{2\phi} &=& F 
\left(\frac{H}{H'}\right)^{2\delta}\nn
F_{[3]} &=& b\,\,{\rm Vol}(H_{3})
\eea
with the parameter relation as given in (3.21). From (3.29) we now find 
$\delta = -3n$ and from (3.30), $F=F_{\rm KMP}(H/H')^n$. Substituting
these in (3.34), we get the isotropic SNS5-brane solution as,
\bea
ds^2 &=& F_{\rm KMP}(HH')\left(\frac{H}{H'}\right)^{-\frac{n}{2}}
\left(-dt^2 + t^2
dH_3^2\right) + \left(\frac{H}{H'}\right)^{-\frac{3n}{2}}
\left(dx_1^2 + \cdots + 
dx_6^2\right)\nn
e^{2\phi} &=& F_{\rm KMP} 
\left(\frac{H}{H'}\right)^{-5n}\nn
F_{[3]} &=& b\,\,{\rm Vol}(H_{3})
\eea
This is precisely the isotropic SNS5-brane solution given in eq.(23) of
ref.\cite{kmp} with $k_2=k_3=k_4=k_5=k_6=-n$, $k_1+\tilde{k}=n$, $k_1-
\tilde{k}=m$. The space-like NS1-brane solution can be obtained from the
SNS5-brane solution of (3.22) and applying the transformations $g_{\mu\nu}
\to g_{\mu\nu}$, $\phi \to -\phi$, $F \to e^{\phi} \ast F$, $q \to d-q$
there (since this is electrically charged) and the solution then matches
exactly with the isotropic SNS1-brane solution given in eq.(20) of \cite{kmp}.
Following a similar procedure M-theory S-branes obtained in \cite{kmp} can 
also be shown to match with the solution (3.22). 

\subsection{Relation with the CGG solution}

S-brane solutions in arbitrary dimensions have also been obtained by CGG in
\cite{cgg}, but they used a different coordinate system from what we have used
in this section. Although CGG solutions have the same symmetry ISO($p+1$)
$\times$ SO($d-p-2,1$) of the metric, their solutions depend only on 
$\hat{t}$, which is the time coordinate of the transverse space and does not
include the other $(d-p-2)$ transverse space-like coordinates. On the other
hand, the solutions we have described depend on the time-like coordinate
$t=(\tilde{t}^2-r^2)^{1/2}$ which includes all the $d-p-1$ transverse 
coordinates. However, since both the solutions have the same symmetry, it
is reasonable to expect that there exists a coordinate transformation by
which these solutions would map to each other. We will show that this is
indeed true. But before we proceed, we should mention that the CGG solutions
differ from the solutions we described in another respect. The CGG solutions 
are characterized by four parameters, whereas our solutions (3.22) are
characterized by three parameters. The origin of this difference is that
unlike in our case, where we assumed that metric becomes flat and $e^{2\phi}
\to 1$ as $t \to \infty$, CGG solutions do not have this property. In order to
compare our solutions with the CGG solutions, we will impose the same 
boundary condition in the CGG solutions and then the latter solutions will also
have three parameters in them and the two solutions will become identical 
under a coordinate transformation.

The general S-brane solutions obtained by CGG are given as \cite{cgg},
\bea
ds^2 &=& \left[\sinh(q-1)\hat{t}\right]^{-\frac{2q}{q-1}}\left[\frac{(d-2)\chi
\tilde{\alpha}^2}{(q-1)b^2}\right]^{-\frac{2(p+1)}{(q-1)\chi}}\left[\cosh
\frac{\chi\tilde{\alpha}}{2}(\hat{t}-t_0)\right]^{\frac{4(p+1)}{(q-1)\chi}}
e^{\frac{2a(p+1)}{(q-1)\chi}(c_1\hat{t}+c_2)}\nn
& &\qquad\qquad\qquad\qquad\qquad\qquad\qquad\qquad\qquad \times 
\left(-d\hat{t}^2 + 
\sinh^2(q-1)\hat{t} dH_q^2\right)\nn
&+& \left[\frac{(d-2)\chi
\tilde{\alpha}^2}{(q-1)b^2}\right]^{\frac{2}{\chi}}\left[\cosh
\frac{\chi\tilde{\alpha}}{2}(\hat{t}-t_0)\right]^{-\frac{4}{\chi}}
e^{-\frac{2a}{\chi}(c_1\hat{t}+c_2)}\left(dx_1^2 + \cdots + dx_{p+1}^2\right)
\\ 
e^{2\phi} &=& \left[\frac{(d-2)\chi
\tilde{\alpha}^2}{(q-1)b^2}\right]^{\frac{2a(d-2)}{(q-1)\chi}}\left[\cosh
\frac{\chi\tilde{\alpha}}{2}(\hat{t}-t_0)\right]^{-\frac{4a(d-2)}{(q-1)\chi}}
e^{[2-\frac{2a^2(d-2)}{(q-1)\chi}](c_1\hat{t}+c_2)}\\
F_{[q]} &=& b\,\,{\rm Vol}(H_q)
\eea
These solutions are characterized by five parameters $\tilde{\alpha}$, $t_0$,
$c_1$, $c_2$, $b$ with a relation between $\tilde{\alpha}$ and $c_1$ of 
the form,
\be
\frac{(p+1)c_1^2}{\chi} + \frac{\chi\tilde{\alpha}^2(d-2)}{2(q-1)}
= q(q-1)
\ee
We will map this solution to (3.22) by a coordinate transformation given as,
\be
\hat{t} = -\frac{1}{q-1} \ln \frac{H}{H'}
\ee
From (3.40) we obtain,
\be
\left[\sinh(q-1)\hat{t}\right]^{-\frac{2q}{q-1}}\left(-d\hat{t}^2 +
\sinh^2(q-1)\hat{t}\,\, dH_q^2\right) = \frac{(HH')^{\frac{2}{q-1}}}
{(2^{\frac{1}{q-1}}
\omega)^2}\left(-dt^2 + t^2 \,dH_q^2\right)
\ee
Now comparing the transverse part of the metric in (3.36) and (3.22) and
also using (3.41) we find that they match if we identify,
\bea
F &=& \cos^2\theta \left(\frac{H}{H'}\right)^\alpha + \sin^2\theta 
\left(\frac{H'}{H}\right)^\beta \,\,=\,\, \cos^2\theta 
e^{-\alpha(q-1)\hat{t}} + \sin^2\theta e^{\beta(q-1)\hat{t}}\nn
&\equiv& \left[\frac{(d-2)\chi
\tilde{\alpha}^2}{(q-1)b^2}\right]^{-\frac{1}{2}}\left[\cosh
\frac{\chi\tilde{\alpha}}{2}(\hat{t}-t_0)\right]
e^{\frac{a}{2}(c_1\hat{t}+c_2)}\left(2^{\frac{1}{q-1}}\omega\right)^{-\frac{
(q-1)\chi}{2(p+1)}}
\eea
Note that with this identification the longitudinal parts of the metric in the
two solutions also match if we rescale the coordinates $x_i$, for $i=1,
\dots,p+1$ in (3.22) by $x_i \to x_i/(2^{1/(q-1)} \omega)^{(q-1)/(p+1)}$.
We relate the parameters in the two solutions from (3.42) as,
\bea
\alpha &=& \frac{1}{2(q-1)}(\chi \tilde{\alpha} - a c_1)\nn
\beta &=& \frac{1}{2(q-1)}(\chi \tilde{\alpha} + a c_1)\nn
\eea
We also get from there,
\bea
\cos^2\theta &=& \frac{1}{2} e^{\frac{\chi\tilde\alpha}{2}t_0 + 
\frac{ac_2}{2}}\left[\frac{(d-2)\chi\tilde{\alpha}^2}{(q-1)b^2}\right]^{-\frac
{1}{2}}\left(2^{\frac{1}{q-1}}\omega\right)^{-\frac{(q-1)\chi}{2(p+1)}}\nn
\sin^2\theta &=& \frac{1}{2} e^{-\frac{\chi\tilde\alpha}{2}t_0 + 
\frac{ac_2}{2}}\left[\frac{(d-2)\chi\tilde{\alpha}^2}{(q-1)b^2}\right]^{-\frac
{1}{2}}\left(2^{\frac{1}{q-1}}\omega\right)^{-\frac{(q-1)\chi}{2(p+1)}}
\eea
From (3.43) we obtain,
\bea
\tilde\alpha &=& \frac{(q-1)}{\chi} (\alpha+\beta)\nn
{\rm and} \quad ac_1 &=& - (q-1)(\alpha-\beta), \qquad {\rm or,} \quad
c_1 \,\,=\,\, -(q-1) \delta
\eea
Using (3.45) the parameter relation (3.39) reduces to
\be
\frac{(p+1)c_1^2}{\chi} + \frac{\chi \tilde{\alpha}^2 (d-2)}{2(q-1)} = q(q-1)
\quad \Rightarrow \quad \frac{1}{2}\delta^2 + \frac{2\alpha(\alpha-a\delta)
(d-2)}{\chi(q-1)} = \frac{q}{q-1}
\ee
This is precisely the parameter relation we obtained in (3.21). On the other
hand from (3.44) we obtain,
\bea
\tan\theta &=& e^{-\frac{\chi\tilde\alpha}{2}t_0}\nn
2^{\frac{1}{q-1}}\omega &=& \frac{e^{\frac{a(p+1)}{(q-1)\chi}c_2} \left[
\cosh\frac{\chi\tilde\alpha}{2}t_0\right]^{\frac{2(p+1)}{(q-1)\chi}}}
{\left[\frac{(d-2)\chi\tilde{\alpha}^2}{(q-1)b^2}\right]^{\frac{p+1}
{(q-1)\chi}}}
\eea
Comparing the dilaton expressions (3.37) and (3.22) we further find that
they match provided,
\be
\left(2^{\frac{1}{q-1}}\omega\right)^{\frac{a(d-2)}{p+1}} = e^{c_2}
\ee
Eliminating $\omega$ in the last relation in (3.47) and (3.48), $c_2$ gets
fixed in terms of other parameters in the CGG solutions as,
\be
c_2 = \frac{a(d-2)}{2(p+1)(q-1)}\ln\left[\frac{(q-1)b^2}{(d-2)\chi
\tilde{\alpha}^2}\cosh^2\frac{\chi\tilde\alpha}{2}t_0\right]
\ee
This shows that if we map the CGG solutions to the solutions (3.22) then one
of the parameters of the CGG solutions gets removed and we are left with
three parameter solutions. The origin of this phenomenon is the fact that 
while mapping the CGG solutions to (3.22) we are imposing the same boundary 
condition ( the metric becoming flat and $e^{2\phi} \to 1$ as $t \to \infty$
) to the CGG solutions and this removes an additional freedom in the
CGG solutions. Thus we have performed a complete mapping of the CGG solutions
described by the parameters $\tilde\alpha$, $c_1$, $t_0$, $b$ (with a
relation between $\tilde\alpha$ and $c_1$) to the solutions obtained in (3.22)
described by the parameters $\alpha$, $\delta$, $\omega$, $\theta$ (with a 
relation between $\alpha$ and $\delta$). The parameters in these two solutions
are related by
\bea
c_1 &=& -(q-1) \delta\nn
\tilde\alpha &=& \frac{(q-1)(2\alpha - a\delta)}{\chi}\nn
t_0 &=& - \frac{2}{(q-1)(2\alpha - a\delta)} \ln\tan\theta\nn
b &=& 4 \omega^{q-1} (2\alpha - a\delta) \sqrt{\frac{(d-2)(q-1)}{\chi}}
\sin\theta \cos\theta
\eea
Note that these relations are exactly the same in $d=10$ as obtained in
eq.(3.17) of ref.\cite{srone}.

\sect{Another non-extremal Euclidean brane solutions}

In this section we will discuss another class of non-extremal Euclidean 
brane (other
that the one discussed in section 3) solutions of the equations of motion
(2.2) -- (2.4). As opposed to the solutions obtained in (3.22), these 
solutions will not be real and we will interpret the corresponding 
real solutions as the non-extremal
E-brane solutions of type II$^\ast$ theories when the form fields are in the
RR sector. In the previous section we found from the consistency of the
equations of motion that the non-extremality function $G(t)$ must be
restricted by the equation (3.9). If we do not assume $G(t) \to 1$ as
$t \to \infty$, then $G(t)$ could take the form,
\be
G(t) = \frac{\omega^{2(q-1)}}{t^{2(q-1)}} = H^2(t)
\ee
where $H(t) = \omega^{q-1}/t^{q-1}$ is a harmonic function in the 
$(q+1)$-dimensional transverse space. One of the motivations for constructing
such solutions was to look at the solutions (3.22) in the $t \to 0$ limit.
It is known that for the static BPS D3-brane $r \to 0$ limit gives rise
to the AdS$_5$ $\times$ S$^5$ solution of type IIB supergravity. Similarly
for the time dependent solutions $t \to 0$ limit can be expected to give 
de Sitter type solution. However, it is clear from the solutions (3.22) that
it is not possible to take $t \to 0$ limit directly as this will make
$H = 1 - \omega^{q-1}/t^{q-1}$ negative and so the solutions will not remain
real. One way to avoid the constant term in $G(t)$ (given in (3.10))
is to consider the solution of eq.(3.9) of the form given in (4.1). However,
we will see that even this form of $G(t)$ does not lead to real solutions
of type II supergravities. 

Now comparing $B$ and $\phi$ equations
(3.5) and (3.8) we obtain,
\be
\phi = \frac{a(d-2)}{q-1} B
\ee
Using the form of $G$ in (4.1) and $\phi$ in (4.2) and assuming $B$ to be
of the form $B=\alpha\ln H$, where $\alpha$ is a parameter to be determined
from the equations of motion, we find that eq.(3.5) reduces to,
\be
\alpha (q-1)^2 \omega^{2(q-1)} + \frac{b^2(q-1)}{2(d-2)} \frac{t^{2(q-1)}
\omega^{\alpha\chi(q-1)}}{t^{\alpha\chi(q-1)} \omega^{2(q-1)}} = 0
\ee
This equation can be solved if $\alpha\chi=2$ and then the solution is
\be
\omega^{2(q-1)} = - \frac{b^2\chi}{4(d-2)(q-1)}
\ee
Note that since $\chi>0$, the harmonic function $H$ will be real only if
$b^2<0$, or, $b$ is purely imaginary. This implies that the field-strengths
are purely imaginary. It can be easily checked that the $R_{tt}$ equation (3.7)
is automatically satisfied with this solution. Now since $\alpha=2/\chi$ we 
have,
\be
A = - \frac{p+1}{q-1} B + \ln H^2 = \ln H^{-\frac{2(p+1)}{(q-1)\chi} + 2}
\ee
and so,
\be
e^{2B} = H^{\frac{4}{\chi}}, \qquad e^{2A} = H^{-\frac{4(p+1)}{(q-1)\chi} + 4}
\ee
The complete solutions therefore are given as,
\bea
ds^2 &=& H^{-\frac{4(p+1)}{\chi(q-1)}+4}\left(-dt^2 + t^2 dH_{d-p-2}^2\right)
+ H^{\frac{4}{\chi}}\left(dx_1^2 + \cdots + dx_{p+1}^2\right)\nn
e^{2\phi} &=& H^{\frac{4a(d-2)}
{(q-1)\chi}}\nn
F_{[q]} &=& b\,\,{\rm Vol}(H_q)
\eea
where $b$ is purely imaginary. These solutions look very similar to the
extremal solutions obtained in eq.(2.20). The powers of the harmonic
functions in these two solutions differ by a sign and there is an additional
$H^4$ factor in front of the transverse part of the metric in (4.7). This
extra $H^4$ factor is due to the presence of the non-extremality function 
$G(t)$ which was absent for the solution (2.20). However, note that the
harmonic functions $H(t)$ in these two solutions are different in general.
They match only in the limit $t \to 0$. Also for the extremal solutions
(2.20), $H(t) \to 1$ for $t \to \infty$ and the metric becomes flat, but 
for the solutions (4.7), $t \to \infty$ limit is ill-defined. 
Eq.(4.7) represent non-extremal
Euclidean branes and as discussed at the end of section 2, when the 
field-strengths belong to the RR sector, we can interpret the real
solutions as E-branes
of type II$^\ast$ theories.

\sect{Conclusion}

To summarize, in this paper we have constructed various time dependent
solutions of supergravity equations of motion containing a dilaton and
a $q$-form field-strength in arbitrary dimensions. We have directly solved
the non-linear equations of motion (unlike the method used in \cite{kmp}) of
the corresponding action. The metrics are assumed to have the symmetries
ISO($p+1$) $\times$ SO($d-p-2,1$) and the field-strengths are assumed to
be magnetic, so, they represent the $(p+1)$-dimensional magnetically charged
Euclidean branes. We found that when the metric components satisfy an
extremality condition, similar to the static BPS $p$-branes, then the
magnetic charges of these solutions become imaginary. But when the 
field-strengths belong to the RR sector, these solutions can become real
if we think of them as the solutions of the so-called `starred' theories
instead of the ordinary type II theories. In that case the solutions would 
represent E$p$-branes of type II$^\ast$ theories. 
We found that this problem does not arise if we relax the extremality
condition and in that case we obtained real time-dependent solutions of
the type II supergravity equations of motion. These solutions are the 
generalizations
of the supergravity S$p$-brane solutions obtained by KMP (in $d=10$) to
arbitrary dimensions. We observed that in order to solve the equations
of motion consistently we need to introduce at least three parameters, whose
physical meanings are not clear to us. These solutions (as pointed out
in \cite{kmp}) have generic singularities at $t=\omega$, whose resolution
is an important open problem to understand. We showed how our solutions
exactly reduce to the solutions found by KMP in $d=10$. S$p$-brane solutions
in a different coordinate systems were also obtained by CGG \cite{cgg} in
arbitrary dimensions. However, the solutions in \cite{cgg} are characterized
by four parameters instead of three as in our case. This difference is
due to the use of different boundary conditions in these two sets of solutions.
When we used the same boundary conditions we showed that the CGG solutions
get mapped exactly to our solutions by a coordinate transformation given
in (3.40). We have 
also given the relations between the parameters in these two solutions.
Finally, we have obtained another class of non-extremal Euclidean brane
solutions. These solutions are not real and as before when the field-strengths
belong to the RR sector, they can be made real by interpreting them as the 
non-BPS E$p$-brane 
solutions of type II$^\ast$ string theories. We pointed out similarities
of these non-extremal solutions and the extremal solutions obtained in
section 2.

\section*{Acknowledgements}

We would like to thank Sudipta Mukherji for discussions.

\end{document}